\documentclass[a4paper,11pt]{article}
\usepackage{pos}

\title{Addressing Selected Gamma-Ray Burst Science Topics with Future Space Instruments}
\ShortTitle{Addressing Selected GRB Science Topics with Future Space Instruments}

\author*[a]{Nicolas De Angelis}

\affiliation[a]{INAF-IAPS,\\
  100 via del Fosso del Cavaliere, Rome, Italy}

\emailAdd{nicolas.deangelis@inaf.it}

\abstract{Gamma-ray bursts (GRBs) are among the most energetic events in the universe, offering insights into stellar collapse, extreme matter behavior, and cosmic evolution. The advent of multi-messenger astronomy, combining electromagnetic, gravitational wave, and neutrino observations, alongside advances in high-energy polarimetry, is revolutionizing GRB research, enabling deeper exploration of their physical mechanisms.

This manuscript summarizes how upcoming and proposed space-based missions will tackle key challenges in GRB science, focusing on four areas: \textit{(i)} identifying high-redshift GRBs to probe the early universe, \textit{(ii)} enhancing multi-messenger detection and localization, \textit{(iii)} improving multi-wavelength follow-up of GRB afterglows, and \textit{(iv)} studying prompt emission polarization to understand jet dynamics and magnetic fields. Highlighting planned missions and their advancements, this work provides a snapshot of current GRB research frontiers, with updates on the evolving status of these missions.}

\FullConference{%
Frontier Research in Astrophysics – IV (FRAPWS2024)\\
9-14 September 2024\\
Mondello, Palermo, Italy
}

\begin{document}
\maketitle

\newpage
\section{Introduction}

Gamma-ray bursts (GRBs) are some of the most energetic and enigmatic events in the universe, with the power to release immense amounts of energy over brief durations. Their study has the potential to shed light on stellar collapse, the behavior of matter under extreme conditions, and even the structure and evolution of the universe. With the recent rise of multi-messenger astronomy, which combines electromagnetic observations with gravitational waves (GWs) and neutrinos, GRB research has entered a new era. Furthermore, the advancement in high-energy polarimetry enlarges the sets of parameters available to study the physical mechanisms at play in extreme environments, with a great disentangling power.\\

This paper explores how upcoming and proposed space-based missions will address key challenges in GRB science across four primary areas: \textit{(i)} identifying high-redshift GRB populations to probe the early universe, \textit{(ii)} improving multi-messenger capabilities for GRB detection and localization, \textit{(iii)} enhancing multi-wavelength follow-up of GRB afterglows, and \textit{(iv)} studying the polarization of the prompt emission of GRBs to understand jet dynamics and magnetic fields structure. Each of these topics represents a frontier of GRB research, and the instruments reviewed here are poised to push these frontiers forward. This is not intended to be an exhaustive list of GRB-related science topics or planned instruments, but rather a compact snapshot of the current landscape of GRB research from space. Examples of proposed and planned missions will be given for each of these fields of research based on the review in \cite{GRB_instr_review}, to which the reader is referred for further readings. The final section will explore the rapidly evolving status of the space missions discussed in this manuscript, offering the most up-to-date information available.

\newpage
\section{High-Redshift GRB Population}  

High-redshift gamma-ray bursts (GRBs) provide a unique observational window into the early universe, enabling the study of the first stars, galaxies, and the cosmic re-ionization epoch. These GRBs, occurring at redshifts \( z > 6 \), trace the earliest stages of stellar evolution and the intergalactic medium (IGM). Upcoming space missions are designed to enhance the detection and characterization of these high-redshift GRBs, allowing astronomers to probe fundamental questions about the formation and evolution of cosmic structures. It should also be noted that recent developments in machine learning techniques have significantly improved redshift inference for GRBs \cite{Dainotti24_ApJS, Dainotti24_ApJL, Narendra25_AandA}, and advances in light-curve reconstruction methods have reduced the number of GRBs required to perform GRB cosmology \cite{Dainotti22_MNRAS, Dainotti23_ApJS}.

\begin{description}
    \item[Gamow Explorer] \cite{Gamow} is specifically optimized for detecting GRBs at \( z > 6 \). It employs a lobster-eye X-ray telescope (LEXT) with a wide field of view (FoV) of 1350 square degrees, allowing efficient detection of GRBs. Once a candidate GRB is detected, its photo-z infrared telescope (PIRT) autonomously slews to refine localization and identify high-redshift events using the photo-z technique. Gamow aims to detect at least 20 high-redshift GRBs during its mission lifetime, providing data critical for understanding re-ionization through the Lyman-$\alpha$ absorption signature and metallicity of host galaxies.
    \item[HiZ-GUNDAM] \cite{HiZ-GUNDAM}, a JAXA mission, focuses on high-redshift GRBs as one of its primary scientific goals. Its wide-field X-ray monitor (WFXM) uses lobster-eye optics to monitor a FoV exceeding 0.5 steradian in the 0.5–4 keV range. Detected GRBs are followed up by the near-infrared telescope (NIRT), which simultaneously observes in five NIR bands to locate and analyze high-redshift GRBs. HiZ-GUNDAM’s ability to observe GRB afterglows for over 10 minutes ensures detailed spectral and photometric data for probing early universe phenomena.
    \item[SVOM (Space-based Variable astronomical Object Monitor)] \cite{SVOM} is tailored to detect and follow up on X-ray-rich and ultra-long GRBs. Its space segment includes ECLAIRs, a coded-mask X-ray telescope with a low-energy threshold of 4 keV, and the Gamma-Ray Monitor (GRM), which extends the spectral range to 5 MeV. Complementing these are the ground-based GWACs (Ground-Based Wide-Angle Cameras) and robotic follow-up telescopes for detailed optical and infrared afterglow studies. By operating with an anti-solar pointing strategy and rapid alert dissemination, SVOM ensures optimized observation conditions for high-redshift GRBs.
    \item[THESEUS (Transient High-Energy Sky and Early Universe Surveyor)] \cite{THESEUS} is a European Space Agency mission concept designed to explore the early universe through the detection of high-redshift GRBs. THESEUS combines a Soft X-ray Imager (SXI) with a Wide Field Monitor (WFM) for efficient detection and localization. The Infrared Telescope (IRT) further aids in photometric and spectroscopic follow-up of high-redshift events. THESEUS is expected to significantly expand the sample of \( z > 6 \) GRBs, enabling breakthroughs in our understanding of the re-ionization epoch and early star formation.
\end{description}

Collectively, these missions will advance our understanding of the high-redshift universe by providing precise measurements of GRB host galaxies, star formation rates, and the chemical enrichment of the early cosmos. The synergy between wide-field detection capabilities and rapid, multi-wavelength follow-up will significantly expand our knowledge of the conditions prevalent during the universe’s formative stages.

\newpage
\section{Multi-Messenger GRB Detection and Localization}  

The integration of gamma-ray bursts into the realm of multi-messenger astronomy represents a transformative advance in astrophysics. GRBs associated with gravitational wave (GW) sources, such as binary neutron star (BNS) and neutron star-black hole (NS-BH) mergers, provide a direct probe of relativistic jets, compact object physics, and the origin of heavy elements via kilonovae. Future missions will enhance the detection and localization of GRBs in the multi-messenger context.

\begin{description}
    \item[MoonBEAM] \cite{MoonBEAM} aims to deliver unparalleled all-sky gamma-ray monitoring from a cislunar orbit, a location free from terrestrial radiation and occultation effects. Its six scintillating detector assemblies provide instantaneous coverage, with localization capabilities akin to the Fermi Gamma-ray Burst Monitor (GBM). This mission’s primary role in multi-messenger astronomy is its capacity to issue rapid alerts for prompt emission counterparts to GW events, enabling follow-up observations by ground and space-based telescopes.
    \item[Gamow Explorer] \cite{Gamow} combines rapid detection with precision localization, making it a cornerstone for multi-messenger studies. Upon receiving a GW alert, Gamow’s lobster-eye X-ray telescope reorients to scan the uncertainty region, identifying GRB counterparts within minutes. Its capability to localize GRBs to 3 arcmin and autonomously follow up with photometric measurements ensures rapid dissemination of accurate data to the broader astronomy community.
    \item[StarBurst] \cite{StarBurst}, a NASA Pioneer-class SmallSat mission, is designed to detect GRBs coincident with GW events. Its NaI(Tl) scintillators, arranged in a half-cube configuration, achieve high detection sensitivity for short GRBs (SGRBs) out to distances of 600 Mpc. Localization accuracies of less than 5 degrees are sufficient to enable follow-up by robotic telescopes and large observatories.
    \item[COSI] (Compton Spectrometer and Imager) \cite{COSI} is a NASA Small Explorer mission scheduled for launch in 2027, which will advance multi-messenger GRB science through its unique capabilities in the 0.2–5 MeV range. As a Compton telescope, COSI will autonomously trigger on GRBs, localize them to $\sim$1$^\circ$ accuracy within an hour, and disseminate rapid alerts to the community. Its high-resolution spectroscopy will enable detailed studies of GRB prompt emission in the MeV band, including spectral features and potential nuclear line signatures. COSI’s operational timeline aligns with the LIGO-Virgo-KAGRA O5 gravitational wave observing run, positioning it as a key player in identifying electromagnetic counterparts to compact binary mergers. The mission’s polarization sensitivity further complements its multi-messenger role, offering potential joint analyses with contemporaneous polarimeters like POLAR-2 (see section~\ref{sec:polarimetry}).
    \item[eXTP's W2C] (Wide Field and Wideband Camera) \cite{new_eXTP} is a proposed replacement for the original Wide Field Monitor, aiming to provide enhanced capabilities for GRB and transient detection over a broader energy range. The instrument features a coded-mask aperture coupled with a segmented GAGG scintillator array, enabling sensitivity across a wide energy band and improving detection efficiency for both short and long GRBs. The W2C design benefits from advanced electronics developed for the POLAR-2 mission, specifically the universal SiPM readout system \cite{POLAR-2_FEE}, which offers high timing resolution, low noise, and flexible triggering modes. This shared technology base ensures proven performance while streamlining development and integration for eXTP.
    \item[STROBE-X] \cite{STROBE-X, STROBE-X_WFM, STROBE-X_HEMA, STROBE-X_LEMA}, with its wide-field monitor (WFM) and pointed instruments, provides complementary capabilities. The WFM will detect and localize short GRBs with arcminute precision, while its rapid slewing ability ensures prompt multi-wavelength follow-up. By focusing on GW sources, STROBE-X bridges the gap between GW detection and electromagnetic (EM) characterization.
    \item[HERMES-Pathfinder] \cite{HERMES} is a constellation of six 3U nano-satellites designed for precise localization of gamma-ray bursts (GRBs) and other high-energy transients. The mission aims to demonstrate the feasibility of using a distributed network of small, cost-effective satellites to enhance GRB detection and localization capabilities. Each satellite carries an innovative X-ray and gamma-ray detector system based on Silicon Drift Detectors (SDDs) coupled with GAGG scintillators, allowing for precise time-domain astrophysics studies. The constellation will leverage the triangulation of signal arrival times to achieve sub-degree localization accuracy, complementing larger missions in the multi-messenger astronomy era.
\end{description}

The collective efforts of these missions will enhance our ability to detect and study GRBs as multi-messenger events. Their design ensures rapid response times, precise localization, and seamless integration into the global multi-messenger network, fostering a deeper understanding of extreme astrophysical phenomena.

\newpage
\section{Multi-Wavelength Afterglow Follow-up}  

The afterglows of gamma-ray bursts provide a rich dataset for understanding the progenitor systems, the surrounding environment, and the physical processes driving the explosions. Multi-wavelength observations of these afterglows are critical for disentangling the complex interplay between jets, magnetic fields, and the circumburst medium. Future missions aim to significantly enhance the coverage and sensitivity of afterglow observations across the electromagnetic spectrum.

\begin{description}
    \item[Einstein Probe] \cite{EP} is designed to monitor the soft X-ray sky with unprecedented sensitivity and cadence, using its wide-field X-ray telescope (WXT) and follow-up X-ray telescope (FXT). The WXT’s large field of view allows it to rapidly detect GRBs, while the FXT provides precise localization and spectral characterization of the afterglow. This capability is essential for studying the transition between the prompt and afterglow phases of GRBs.
    \item[SVOM] \cite{SVOM} combines space- and ground-based instruments for comprehensive multi-wavelength afterglow follow-up. The satellite’s ECLAIRs telescope detects GRBs and provides localization, while its Microchannel X-ray Telescope (MXT) and Visible Telescope (VT) perform detailed afterglow studies in X-ray and optical wavelengths, respectively. On the ground, the GWACs and robotic follow-up telescopes add further depth to the optical and infrared follow-up efforts, ensuring rapid and accurate characterization of GRB afterglows.
    \item[HiZ-GUNDAM] \cite{HiZ-GUNDAM} offers unique capabilities for multi-wavelength follow-up with its dual payload of the WFXM and NIRT. The WFXM’s wide field of view ensures efficient detection of GRBs, while the NIRT’s near-infrared observations provide crucial data for understanding the high-redshift afterglow population and their host environments. This mission is particularly valuable for investigating GRBs in the early universe.
    \item[STROBE-X] \cite{STROBE-X, STROBE-X_WFM, STROBE-X_HEMA, STROBE-X_LEMA} contributes to afterglow follow-up with its high-energy monitoring capabilities. Its WFM detects GRBs and localizes them within minutes, enabling rapid transition to pointed observations with its narrow-field instruments. These observations provide high-resolution temporal and spectral data critical for modeling GRB afterglows and understanding their central engines.
    \item[THESEUS] \cite{THESEUS} stands out with its dual capability to detect GRBs and perform multi-wavelength follow-up. Its Soft X-ray Imager (SXI) and Infrared Telescope (IRT) are particularly suited to follow up on GRB afterglows across X-ray and infrared bands, offering unique insights into the interplay between the jet and its environment. The mission's capabilities will provide detailed temporal evolution and multi-wavelength characterization of GRB afterglows.
\end{description}

The coordinated efforts of these missions will enable unprecedented multi-wavelength coverage of GRB afterglows, providing insights into their progenitor systems, energy release mechanisms, and interaction with the surrounding medium. Such observations will deepen our understanding of the extreme physics governing these powerful cosmic events.

\newpage
\section{Prompt Emission Polarimetry}\label{sec:polarimetry}  

Polarimetry of the prompt emission in gamma-ray bursts holds the potential to unveil the geometry and magnetic field structure of GRB jets, offering a direct probe into their launching mechanisms and energy dissipation processes. Observing the degree and angle of polarization in prompt emission provides insights into the jet composition and the role of magnetic fields in shaping the observed radiation. Some proposed GRB-dedicated polarimeters are:

\begin{description}
    \item[POLAR-2] \cite{POLAR-2, NDA_thesis}, scheduled for a launch to the China Space Station in 2027, represents a significant advancement in GRB polarimetry. With a detector array four times larger than its predecessor, POLAR \cite{POLAR_instr, POLAR_catalog}, this mission will achieve higher sensitivity (especially at tens of keV) and a lower minimal detectable polarization (MDP). POLAR-2 is expected to provide polarization measurements for GRBs with fluences as low as \(10^{-6}\) erg cm\(^{-2}\), allowing detailed studies of jet dynamics and magnetic field configurations. Its rapid alert system will also enable follow-up observations by other facilities.
    \item[LEAP (LargE Area burst Polarimeter)] \cite{LEAP, LEAP2} is a proposed ISS payload that will measure polarization in the 50~keV to 1 MeV~range. As a dual-phase Compton polarimeter, LEAP will distinguish between competing GRB jet models by observing the azimuthal scattering angle distribution of incoming photons. Its large field of view and high sensitivity ensure it can detect and characterize the polarization of numerous GRBs during its mission, providing critical data on jet composition and emission mechanisms. 
    \item[COSI] \cite{COSI} will contribute to GRB polarimetry as a Compton telescope capable of measuring polarization in the 0.2–5 MeV range. While not primarily a dedicated polarimeter, its design allows for polarization studies of bright GRBs, providing independent constraints on jet geometry and magnetic field configurations. The overlap of COSI’s operational period with POLAR-2 (launching in 2027) creates a unique opportunity for cross-instrument polarization analyses, enhancing the robustness of measurements and enabling comparative studies of GRB prompt emission mechanisms.
    \item[eXTP]: The enhanced X-ray Timing and Polarimetry mission (eXTP) \cite{eXTP, new_eXTP} could also contribute to GRB polarization studies, although it is not primarily focused on GRBs. The mission's Polarimetry Focusing Array (PFA) will provide polarimetric capabilities in the 2-8 keV range, complementing the polarization measurements from other missions. While eXTP's primary science goals are centered on neutron stars and black holes, the telescope could be pointed to particularly bright events, as it has been done for the BOAT GRB221009A with IXPE \cite{Negro-BOAT}. 
\end{description}

The systematic study of GRB prompt emission polarimetry by these missions are aimed to resolve long-standing debates about the role of magnetic fields in jet formation and energy dissipation. By correlating polarization data with spectral and temporal observations, these missions will provide a comprehensive picture of the physical processes driving GRBs, marking a significant leap forward in high-energy astrophysics.

\newpage
\section{Mission Status \& Outlook}

Gamma-ray bursts (GRBs) remain one of the most fascinating phenomena in astrophysics, providing key insights into both high-energy astrophysics and fundamental physics. Their role as multi-messenger sources and probes of the early universe has led to a surge in interest from the scientific community, with numerous proposed and ongoing space missions dedicated to their study.

A significant number of missions are in various stages of development, ranging from proposals to operational observatories. Gamow Explorer has been proposed as a NASA MIDEX-class mission, with a potential launch in 2028. HiZ-GUNDAM remains under development, aiming to enhance high-redshift GRB detections. SVOM, launched in 2021, is actively contributing data on X-ray-rich and ultra-long GRBs, while the Einstein Probe, launched in 2024, is already providing promising early results. THESEUS has been proposed as a European Space Agency (ESA) M-class mission with a planned launch in 2037.

Several promising missions remain under active development. MoonBEAM is still in progress, focusing on all-sky gamma-ray monitoring, while StarBurst is expected to begin its one-year nominal science operations in 2027, targeting short GRBs. COSI is on track for launch in 2027 as NASA’s next Small Explorer mission. STROBE-X, which was not selected in the latest NASA Explorers Program call, is likely to be resubmitted in future funding opportunities. The eXTP mission, originally a collaboration between Europe and China, has had its European contribution frozen after phase B. However, a revised design featuring a reduced number of SFA and PFA telescopes is under consideration, with the potential addition of a Wide-Field Wide-Band Camera (W2C), a segmented GAGG detector with a wider energy range than the original Wide Field Monitor (WFM).

Polarimetry remains an exciting frontier in GRB science. The LEAP mission was unfortunately canceled, but efforts are underway to resubmit a modified version for NASA’s Pioneer-class mission call. In contrast, POLAR-2 is advancing steadily, with a planned launch to the China Space Station (CSS) in mid-2027. Meanwhile, the HERMES-Pathfinder constellation, consisting of six CubeSats, is scheduled for launch on March 6, 2025. This pathfinder mission will test the feasibility of a larger future CubeSat-based constellation for GRB detection and localization.

While some missions face cancellations or delays, the continued interest in GRB science ensures a vibrant future for the field. Several other promising missions and proposals not discussed here further highlight the dynamic landscape of GRB research. The growing number of proposed and approved missions demonstrates the strong commitment of the community to addressing fundamental questions about GRBs, their progenitors, and their role in the high-energy universe. The next decade will likely be transformative, with new technologies and mission concepts paving the way for groundbreaking discoveries in GRB astrophysics.

\newpage
\noindent {\bf DISCUSSION}

\bigskip
\noindent {\bf DHEERAJ PASHAM:} How many of these missions are sure to happen? Which one are those?

\bigskip
\noindent {\bf NICOLAS DE ANGELIS:} An update on the status of the various missions can be found the last section.

\bigskip
\noindent {\bf SILVIA ZANE:} What is the status for developing a full constellation mission, following the HERMES pathfinder? 
\bigskip

\textbf{}
\noindent {\bf NICOLAS DE ANGELIS:} As discussed in the last section, the HERMES pathfinder constellation, made of 6 3U-cubesats, has been integrated and will be launched on the 6th of March 2025. A bigger constellation is foreseen in the future, but not yet approved.


\bigskip



\begin{thebibliography}{99}  
\bibitem{GRB_instr_review}
BOZZO, Enrico, et al. Future Perspectives for Gamma-ray Burst Detection from Space. Universe, 2024, 10.4: 187.

\bibitem{Dainotti24_ApJS}
DAINOTTI, Maria Giovanna, et al. Inferring the redshift of more than 150 GRBs with a machine-learning ensemble model. The Astrophysical Journal Supplement Series, 2024, 271.1: 22.

\bibitem{Dainotti24_ApJL}
DAINOTTI, Maria Giovanna, et al. Gamma-ray bursts as distance indicators by a statistical learning approach. The Astrophysical Journal Letters, 2024, 967.2: L30.

\bibitem{Narendra25_AandA}
NARENDRA, Aditya, et al. Gamma-ray burst redshift estimation using machine learning and the associated web app. Astronomy \& Astrophysics, 2025, 698: A92.

\bibitem{Dainotti22_MNRAS}
DAINOTTI, Maria Giovanna, et al. Optical and X-ray GRB Fundamental Planes as cosmological distance indicators. Monthly Notices of the Royal Astronomical Society, 2022, 514.2: 1828-1856.

\bibitem{Dainotti23_ApJS}
DAINOTTI, Maria G., et al. A stochastic approach to reconstruct gamma-ray-burst light curves. The Astrophysical Journal Supplement Series, 2023, 267.2: 42.

\bibitem{Gamow}
WHITE, Nicholas E., et al. The Gamow Explorer: a Gamma-Ray Burst Observatory to study the high redshift universe and enable multi-messenger astrophysics. In: UV, X-Ray, and Gamma-Ray Space Instrumentation for Astronomy XXII. SPIE, 2021. p. 87-100.

\bibitem{HiZ-GUNDAM}
YONETOKU, Daisuke, et al. High-redshift gamma-ray burst for unraveling the dark ages Mission: HiZ-GUNDAM. In: Space Telescopes and Instrumentation 2022: Ultraviolet to Gamma Ray. SPIE, 2022. p. 121811J.

\bibitem{SVOM}
BERNARDINI, Maria Grazia; CORDIER, Bertrand; WEI, Jianyan. The svom mission. Galaxies, 2021, 9.4: 113.

\bibitem{THESEUS}
STRATTA, Giulia, et al. Breakthrough Multi-Messenger Astrophysics with the THESEUS Space Mission. Galaxies, 2022, 10.3: 60.

\bibitem{MoonBEAM}
FLETCHER, Corinne, et al. The Scientific Performance of the MoonBurst Energetics All-sky Monitor (MoonBEAM). arXiv preprint arXiv:2308.16293, 2023.

\bibitem{StarBurst}
KOCEVSKI, Daniel. The StarBurst Multimessenger Pioneer. In: American Astronomical Society Meeting\# 240. 2022. p. 409.07.

\bibitem{COSI}
TOMSICK, John A., et al. The Compton spectrometer and imager. arXiv preprint arXiv:2308.12362, 2023.

\bibitem{new_eXTP}
ZHANG, Shuang-Nan, et al. The enhanced X-ray Timing and Polarimetry mission--eXTP for launch in 2030. arXiv preprint arXiv:2506.08101, 2025.

\bibitem{POLAR-2_FEE}
KOLE, Merlin, et al. Design and performance of a universal SiPM readout system for X-and gamma-ray missions. Nuclear Instruments and Methods in Physics Research Section A: Accelerators, Spectrometers, Detectors and Associated Equipment, 2025, 170782.

\bibitem{STROBE-X}
RAY, Paul S., et al. STROBE-X mission overview. arXiv preprint arXiv:2410.08342, 2024.

\bibitem{STROBE-X_WFM}
REMILLARD, Ronald A., et al. The STROBE-X wide field monitor instrument. Journal of Astronomical Telescopes, Instruments, and Systems, 2024, 10.4: 042505-042505.

\bibitem{STROBE-X_HEMA}
HUTCHESON, Anthony L., et al. Spectroscopic Time-Resolving Observatory for Broadband Energy X-ray high-energy modular array. Journal of Astronomical Telescopes, Instruments, and Systems, 2024, 10.4: 042503-042503.

\bibitem{STROBE-X_LEMA}
GENDREAU, Keith C., et al. STROBE-X low-energy modular array (LEMA) instrument. Journal of Astronomical Telescopes, Instruments, and Systems, 2024, 10.4: 042506-042506.

\bibitem{HERMES}
FIORE, Fabrizio, et al. HERMES-Pathfinder. In: Handbook of X-ray and Gamma-ray Astrophysics. Singapore: Springer Nature Singapore, 2024. p. 1231-1248.

\bibitem{EP}
YUAN, Weimin, et al. The Einstein Probe Mission. In: Handbook of X-ray and Gamma-ray Astrophysics. Singapore: Springer Nature Singapore, 2024. p. 1171-1200.

\bibitem{POLAR-2}
DE ANGELIS, Nicolas, et al. Development and science perspectives of the POLAR-2 instrument: a large scale GRB polarimeter. arXiv preprint arXiv:2109.02978, 2021.

\bibitem{NDA_thesis}
DE ANGELIS, Nicolas. Development of the Next Generation Space-Based Compton Polarimeter and Energy Resolved Polarization Analysis of Gamma-ray Bursts Prompt Emission. 2023. PhD Thesis. University of Geneva.

\bibitem{POLAR_instr}
PRODUIT, Nicolas, et al. Design and construction of the POLAR detector. Nuclear Instruments and Methods in Physics Research Section A: Accelerators, Spectrometers, Detectors and Associated Equipment, 2018, 877: 259-268.

\bibitem{POLAR_catalog}
KOLE, Merlin, et al. The POLAR gamma-ray burst polarization catalog. Astronomy \& Astrophysics, 2020, 644: A124.

\bibitem{LEAP}
MCCONNELL, Mark L., et al. The LargE Area burst Polarimeter (LEAP) a NASA mission of opportunity for the ISS. In: Uv, x-ray, and gamma-ray space instrumentation for astronomy xxii. SPIE, 2021. p. 237-250.

\bibitem{LEAP2}
MELECIO, Karla Oñate, et al. Evaluation of a prototype detector for the LargE Area burst Polarimeter (LEAP). In: UV, X-Ray, and Gamma-Ray Space Instrumentation for Astronomy XXII. SPIE, 2021. p. 251-260.

\bibitem{eXTP}
SANTANGELO, Andrea, et al. The Enhanced X-ray Timing and Polarimetry Mission: eXTP. In: Handbook of X-ray and Gamma-ray Astrophysics. Singapore: Springer Nature Singapore, 2022. p. 1-29.

\bibitem{Negro-BOAT}
NEGRO, Michela, et al. The IXPE view of GRB 221009A. The Astrophysical journal letters, 2023, 946.1: L21.



\end{thebibliography}
\end{document}